\documentclass[%
superscriptaddress,
twocolumn,
amsmath,amssymb,
 preprint
aps,
 prl,
]{revtex4-1}

\usepackage{graphicx}
\usepackage{dcolumn}
\usepackage{bm}
\usepackage{here}
\usepackage{color}
\usepackage{ulem}
\usepackage{textcomp}
\usepackage{longtable}

\begin{document}

\preprint{APS/123-QED}

\title{Distinctive Doping Dependence of Upper Critical Field in Iron-Based Superconductor LaFeAsO$_{1-x}$H$_{x}$}
\author{Shiro Kawachi}
\email{kawachi@sci.u-hyogo.ac.jp}
\affiliation{Materials Research Center for Element Strategy, Tokyo Institute of Technology, Yokohama, Kanagawa 226-8503, Japan.}
\affiliation{Graduate School of Science, University of Hyogo, Koto, Hyogo 678-1297, Japan.}
\author{Jun-ichi~Yamaura}
\email{jyamaura@issp.u-tokyo.ac.jp}
\altaffiliation{The Institute for Solid State Physics, The University of Tokyo at present}
\affiliation{Materials Research Center for Element Strategy, Tokyo Institute of Technology, Yokohama, Kanagawa 226-8503, Japan.}
\author{Yoshio Kuramoto}
\affiliation{Institute of Materials Structure Science, High Energy Accelerator Research Organization (KEK), Tsukuba, Ibaraki 305-0801, Japan.}
\author{Soshi~Iimura}
\affiliation{Materials Research Center for Element Strategy, Tokyo Institute of Technology, Yokohama, Kanagawa 226-8503, Japan.}
\affiliation{National Institute for Materials Science, 1-1 Namiki, Tsukuba, Ibaraki 305-0044, Japan.}
\author{Toshihiro Nomura}
\affiliation{The Institute for Solid State Physics, The University of Tokyo, Kashiwa, Chiba 277-8581, Japan.}
\author{Yoshimitsu Kohama}
\affiliation{The Institute for Solid State Physics, The University of Tokyo, Kashiwa, Chiba 277-8581, Japan.}
\author{Takashi Sasaki}
\affiliation{Materials Research Center for Element Strategy, Tokyo Institute of Technology, Yokohama, Kanagawa 226-8503, Japan.}
\author{Masashi Tokunaga}
\affiliation{The Institute for Solid State Physics, The University of Tokyo, Kashiwa, Chiba 277-8581, Japan.}
\author{Youichi Murakami}
\affiliation{Institute of Materials Structure Science, High Energy Accelerator Research Organization (KEK), Tsukuba, Ibaraki 305-0801, Japan.}
\author{Hideo Hosono}
\affiliation{Materials Research Center for Element Strategy, Tokyo Institute of Technology, Yokohama, Kanagawa 226-8503, Japan.}
\affiliation{National Institute for Materials Science, 1-1 Namiki, Tsukuba, Ibaraki 305-0044, Japan.}

\begin{abstract}
High magnetic fields up to 105 T have been utilized in deriving the upper critical field $B_{\rm c2}$ of LaFeAsO$_{1-x}$H$_x$ throughout whole temperatures below $T_{\rm c}$.
Resistivity measurements demonstrate that $B_{\rm c2}$ behaves differently in samples with $x = 0.12$ (SC1) from those with 0.32 (SC2).
In SC1, the two-band model assuming the $s$-wave pairing gives a good fitting with repulsive intraband interaction and dominant interband coupling. 
In SC2, we have to assume attractive intraband interaction with weak interband coupling, which in fact suggests a non-$s$-wave pairing in view of the strong Coulomb repulsion.
These results support the possibility that SC1 and SC2 have different pairing symmetries.
\end{abstract}

\maketitle

\par
Iron-based superconductors with high-critical temperature, $T_{\rm c}$, have sparked substantial studies in chemistry and physics since their discovery \cite{Kami2008, Paglione2010, Stewart2011, Hosono2015, Hosono2018}.
Almost all known compounds comprise alternating conduction layers of FeX$_4$ (X = pnictogen, chalcogen) units and charge reservoir layers \cite {Stewart2011}. 
The electronic structure near the Fermi level is dominated by Fe-$3d$ multiorbitals, in contrast to cuprates with single orbital character \cite{Mazin2008,Hosono2015}. 
Both of these superconductors have drawn interest as unconventional superconductors driven by Coulomb repulsion between electrons. 
In iron-based superconductors, the extended $s$-wave model has been considered as critical for Cooper pair formation, in which interband coupling is crucial \cite{Mazin2008, Kuroki2008, Kontani2010}. 
\par
The hydrogen-substituted iron-based superconductor LaFeAsO$_{1-x}$H$_x$ shows a rich phase diagram via electron doping with two antiferromagnetic phases (AF1 and AF2) and two superconducting phases (SC1 and SC2), as seen in Fig.~1 \cite{Iimura2012, Hiraishi2014}.
The magnetic moment 0.63~$\mu_{\rm B}$ in AF1 almost doubles in AF2, and the two magnetic structures show distinct difference (inset of Fig.~1).
The crystal changes from a tetragonal structure to a centrosymmetric- (non-centrosymmetric-) orthorhombic structure in AF1 (AF2) with decreasing temperature \cite{Cruz2008, Qureshi2010, Hiraishi2014}.
Because AF1 and AF2 are different parent phases near SC1 and SC2, respectively, each of superconductivity might have different pairing mechanisms \cite{Yamakawa2013, Iimura2017, Kobayashi2016, Moon2016}.
%
\begin{figure}[t]
 \centering
 \includegraphics[width=7.0cm]{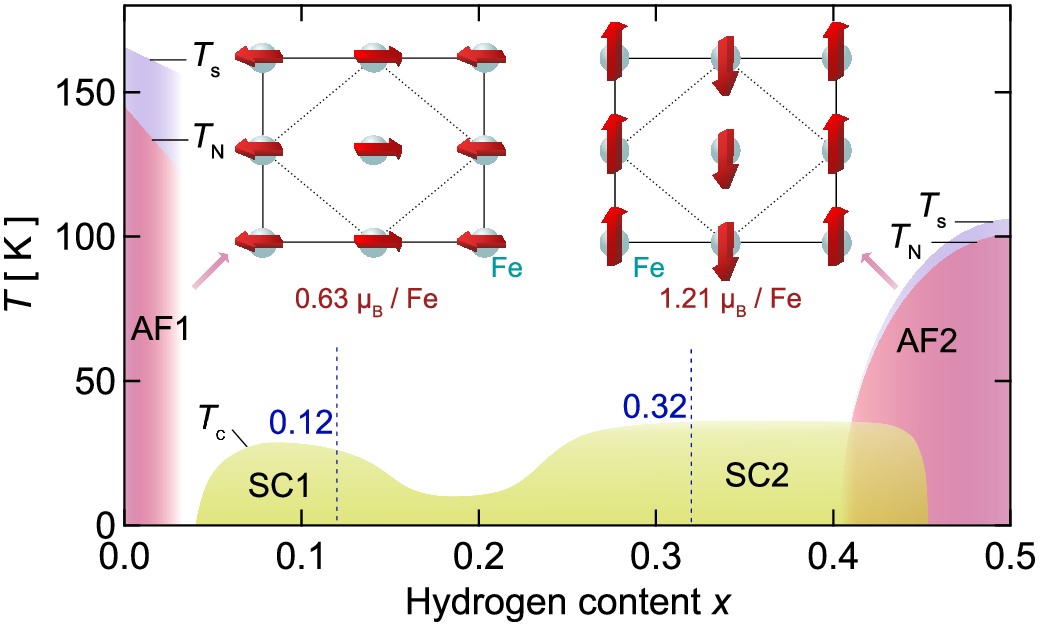}
 \caption{Superconducting, magnetic, and structural phase diagram of LaFeAsO$_{1-x}$H$_x$ with electron doping by hydrogen substitution $x$.
 See text for more detailed information. Temperatures $T_{\rm s}$, $T_{\rm N}$ and $T_{\rm c}$ indicate structural, N\'{e}el and superconducting transition temperatures, respectively. The dotted lines represent the hydrogen content of the samples with $x = 0.12$ and $0.32$. Insets illustrate the magnetic structures in AF1 and AF2 \cite{Qureshi2010,Hiraishi2014}.
}\label{fig:fig1}
\end{figure}
\par
In this work, we investigate the nature of superconductivity in LaFeAsO$_{1-x}$H$_x$ by measuring the upper critical field, $B_{\rm c2}$.
The use of ultra-high magnetic fields over 100 T enabled coverage of the whole temperature ($T$) range below $T_{\rm c}$.
The detailed analysis of $B_{\rm c2}(T)$ may serve to detect different pairing mechanisms depending on the doping.
The temperature dependence of $B_{\rm c2}$ in iron-based superconductors has been studied using Gurevich's two-band theory, which assumes the dirty limit in the $s$-wave model \cite{Gurevich2003, Gurevich2004, Jaroszynski2008, Fuchs2008, Kohama2009, Gurevich2011, Tarantini2011}.
In contrast to prior studies of $B_{\rm c2}$ for iron-based superconductors \cite{Hunte2008, Jaroszynski2008, Kohama2009, Gurevich2011}, we address both attractive and repulsive interactions for intraband couplings.
In SC1, the fitting suggests the two possibilities: (i) repulsive intraband interaction and dominant interband coupling; (ii) attractive intraband interaction and subsidiary interband coupling. 
We propose the possibility (i) because of the strong Coulomb repulsion which is not explicit in the two-band model.
On the other hand, the SC2 requires the intraband attraction and weak interband coupling.  We interpret the result as suggesting the non-$s$-wave pairing since the intraband attraction in the $s$-wave channel is unlikely in view of the strong Coulomb repulsion. 
Moreover, we find that the pair breaking at $B_{\rm c2}$ comes from the orbital effect in SC1, while from the Pauli paramagnetic effect in SC2.
%
\begin{figure}[t]
 \centering
 \includegraphics[width=8.6cm]{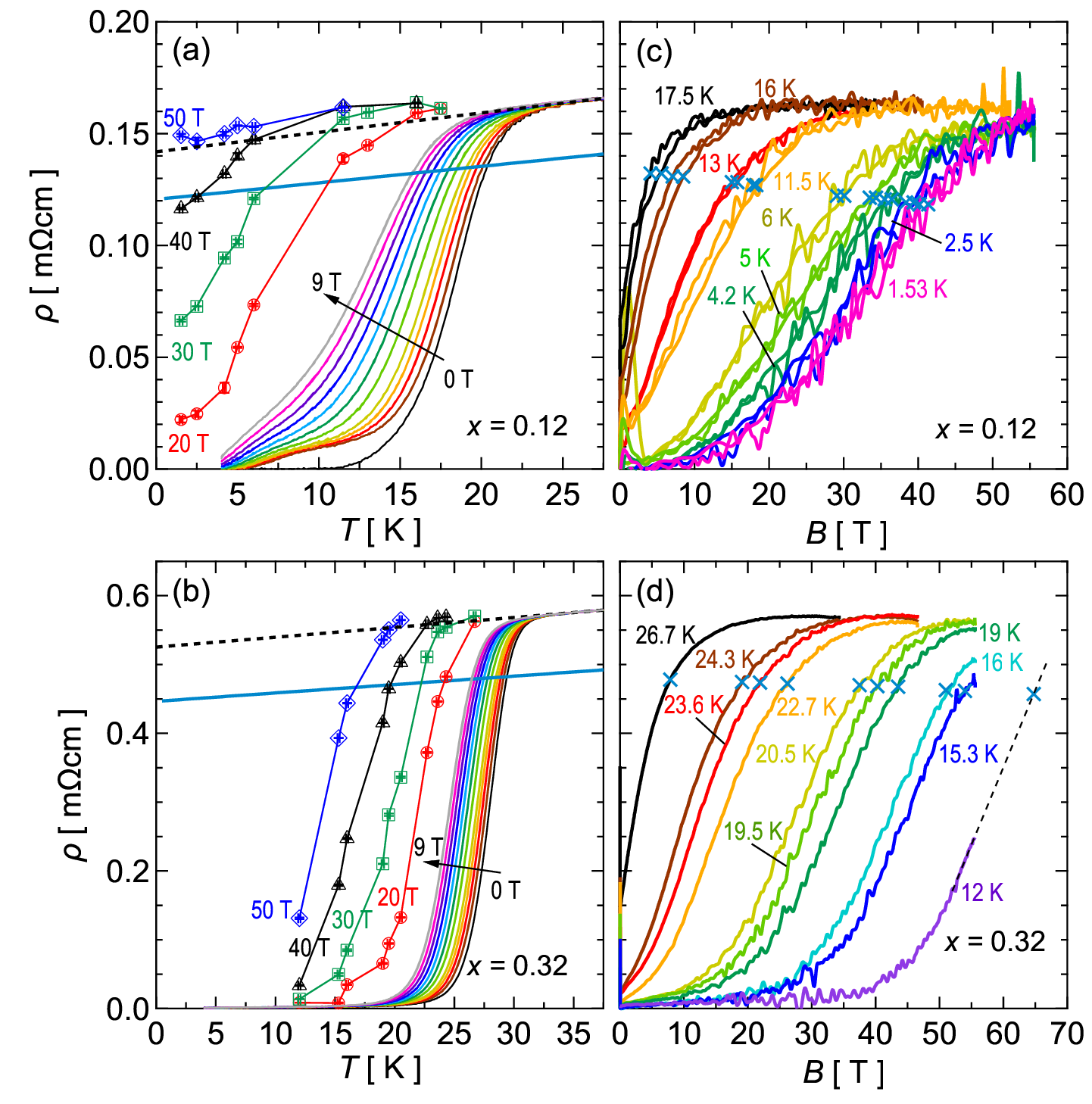}
 \caption{Temperature dependence of electrical resistivity up to 9 T (solid lines) for $x = 0.12$ (a) and 0.32 (b), along with results (symbols) taken from the data described in the text. The magnitudes of applied magnetic field, as indicated by solid lines, are 0, 0.5, 1, 1.5, 2, 3, 4, 5, 6, 7, 8, and 9 T, respectively.
The dashed line represents the extrapolated normal state resistivity, $\rho_{\rm n}$. The upper critical field $B_{\rm c2}$ is defined by the intersection with the 85\% value of $\rho_{\rm n}$ (sky-blue solid line). Magnetic field dependence of the electrical resistivity up to 56 T is shown for $x = 0.12$ (c) and 0.32 (d), where $B_{\rm c2}$ is shown by the crosses (sky blue).
 }\label{fig:fig2}
\end{figure}
\par
We prepared polycrystalline samples of LaFeAsO$_{1-x}$H$_x$ for $x = 0.12$ (SC1) and $0.32$ (SC2) from the solid-state reaction under high-pressure as described in the literature \cite{Iimura2012}.
We employed three different methods to measure the electrical transport according to the magnitude of magnetic fields as follows.
In method-1, up to 9 T, the DC electrical resistance was measured using Physical Property Measurement System (Quantum Design Inc.).
In method-2, up to 56 T, AC electrical resistance was measured at 50 kHz using a non-destructive pulse magnet with a duration of 36 ms.
The same sample was utilized in both kinds of measurements.
In method-3, for the $x = 0.32$ sample alone, radio-frequency (RF) impedance of the sample was measured up to 105 T using a destructive vertical single-turn-coil with the duration of $\sim$6 {\textmu}s \cite{Imamura2006, KawachiUP}.
The sample was connected at the end of the printed circuit with the characteristic impedance of 50 $\Omega$. 
The reflection amplitude and phase of the RF signal altered depending on the impedance matching at the sample. 
We measured the reflectance at 2 and 4.2 K and converted to the electrical resistivity \cite{SM}. 
For method-3, we utilized a sample from the same batch as used in methods 1 and 2. 
\par
Figures 2(a) and 2(b) show the temperature dependence of the electrical resistivity, $\rho$, up to 9 T for $x=0.12$ and 0.32, respectively.
Upon cooling the resistivity drops sharply at the superconducting transition.
The data coincide with those upon heating across $T_c$.
We estimate the normal state resistivity, $\rho_{\rm n}$, from linear extrapolation of the data as depicted by the dashed line. 
Since $\rho_{\rm n}$ has almost no magnetoresistance in both compounds, we define the upper critical field, $B_{\rm c2}$, by the intersection with the 85\% value of $\rho_{\rm n}$ (sky-blue solid line). 
The criterion of 85\% of $\rho_{\rm n}$ was adopted to estimate $B_{\rm c2}$ from as many experimental data as available on the high-resistance side. 
\par
Figures 2(c) and 2(d) show the magnetic-field dependence of $\rho$ up to 56 T for $x=0.12$ and 0.32.
The $\rho$ grows sharply with the breaking of superconductivity for both compounds.
The crosses (sky blue) indicate $B_{\rm c2}$ at 85\% of $\rho_{\rm n}$ determined by method-1.
The value of $B_{\rm c2}$ at 12 K for $x = 0.32$ is estimated using linear extrapolation of 52--56~T data because
the resistivity does not yet reach 85\% of $\rho_{\rm n}$.
We extracted the temperature dependence for $B>9$ T from the data taken by method-2, as shown  in Figs.~2(a) and 2(b).
This procedure allowed us to combine the data obtained by methods 1 and 2. 
%
\begin{figure}[t]
 \centering
 \includegraphics[width=7.0cm]{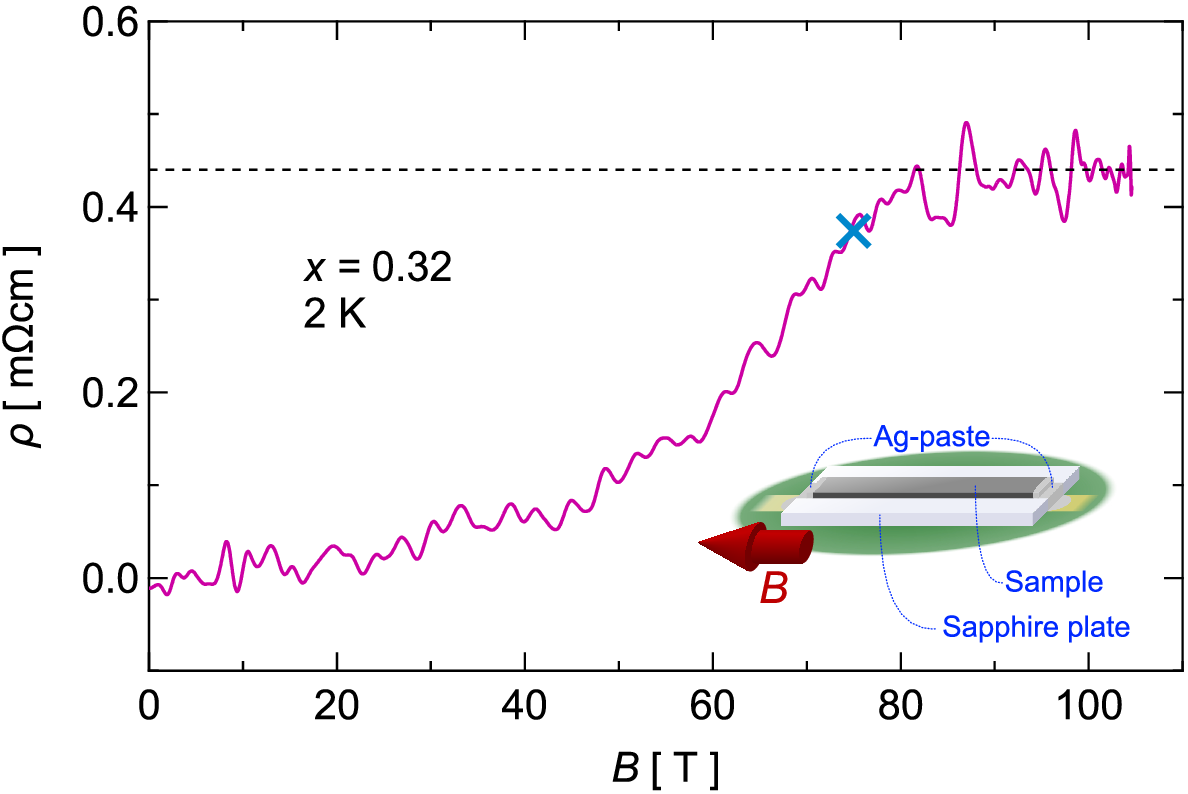}
 \caption{
Magnetic field dependence of electrical resistivity, $\rho$, at 2 K up to 105 T for $x=0.32$. The cross (in blue) represents $B_{\rm c2}$ at 85\% of the estimated normal state resistivity (dashed line).
 }\label{fig:fig3}
\end{figure}
\par
Figure 3 plots the magnetic field dependence of the electrical resistivity at 2 K up to 105 T for $x=0.32$.
The resistivity increases gradually upon application of magnetic field, becoming almost constant above 90~T. 
The saturation of $\rho$ (dotted line) over 90~T coincides well with $\rho_{\rm n}$ obtained by method-1. 
Hence, it provides another support to determine $B_{\rm c2}$ based on the 85 \% value of $\rho_{\rm n}$ at 2 and 4.2 K \cite{SM}.
%
\begin{figure}[t]
\label{fig:fig4}
 \centering
 \includegraphics[width=8.0cm]{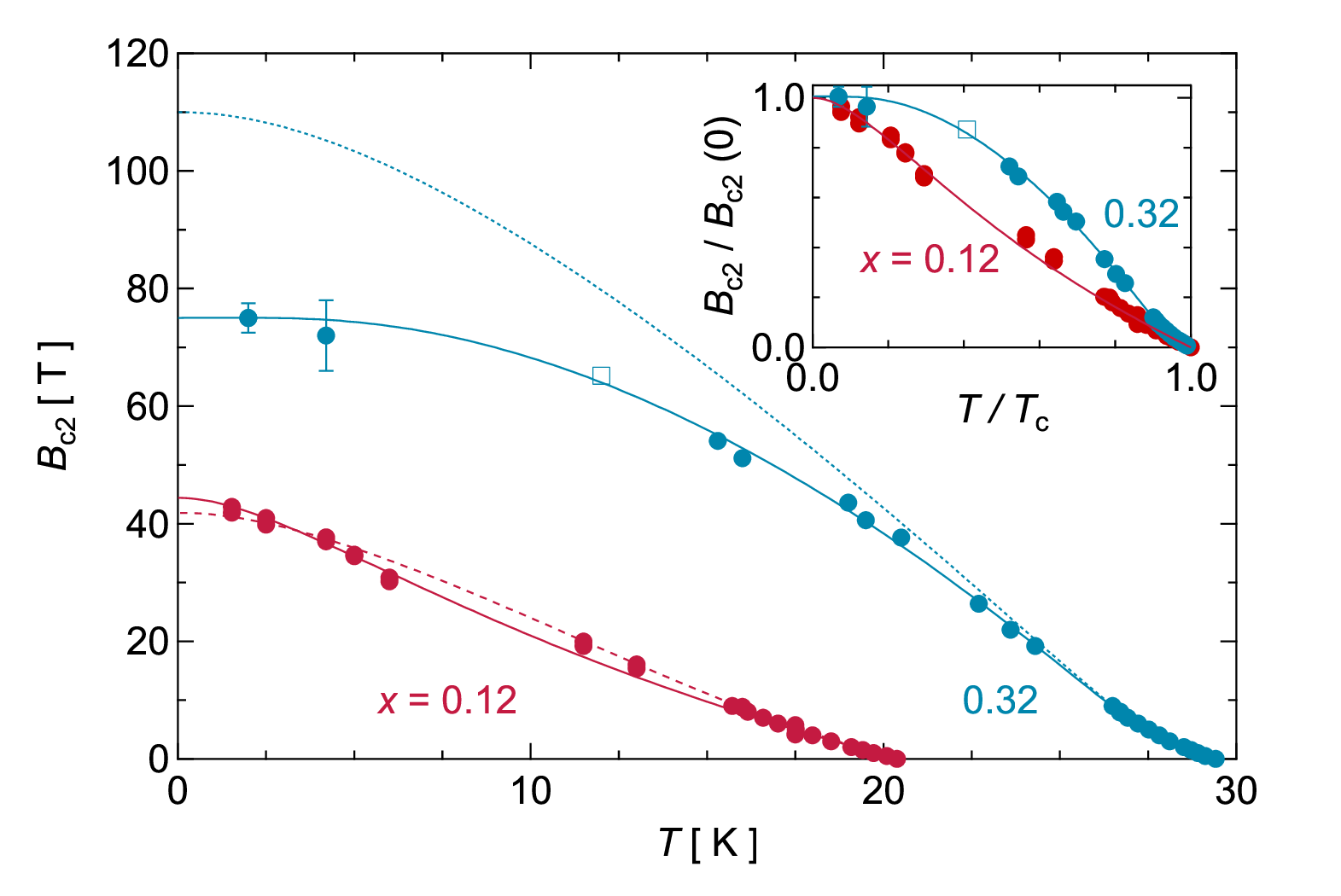}
 \caption{
 Temperature dependence of the upper critical field $B_{\rm c2}$ for $x = 0.12$ and 0.32. 
 A square symbol represents the estimated value extrapolated linearly from the data at 12 K for $x = 0.32$. 
 The error bars take into account reading errors of resistance value. 
For $x = 0.12$, the red solid (dashed) line stands for the fitting curves at $\lambda_{11}=-1$~($\lambda_{11}=1$) with the band index 1 meaning holes.
For $x = 0.32$, blue dotted and solid lines show the results from two-band theory with and without Pauli-pair-breaking effect, respectively. Inset represents the normalized data $B_{\rm c2}(T)/B_{\rm c2}(0)$ vs $T/T_{\rm c}$. The critical temperature $T_{\rm c}$ is estimated at 20.4~K for $x = 0.12$, and at 29.4~K for $x = 0.32$.
 }
\end{figure}
%
\par
Figure 4 plots the $B_{\rm c2}$-$T$ phase diagram (solid circles). 
Large anisotropy in $B_{\rm c2}$ is common in layer-type superconductors \cite{Jaroszynski2008, Kohama2009, Mizuguchi2014}.
$B_{\rm c2}$ is the component corresponding to $B$$||$$ab$ grains on the high-resistance side with the 85\% criteria because the magnetic field initially breaks the superconducting current path in $B$$||$$c$ grains \cite{Mizuguchi2014}. 
For $x = 0.12$, $B_{\rm c2}(T)$ behaves rather linearly in the entire temperature range below $T_{\rm c}$, and reaches $B_{\rm c2}(0) \sim $ 43~T.
This property matches that of the LaFeAsO$_{1-x}$F$_x$ polycrystalline samples, which have the electronic phase diagram similar to the present compound for $x = 0.12$ \cite{Hunte2008, Kohama2009}.
In contrast, for $x = 0.32$, $B_{\rm c2}$ exhibits a concave upward behavior near $T_{\rm c}$, gradually leveling off before reaching $B_{\rm c2}(0) \sim $ 77~T. 
This concave behavior is related to the strong paramagnetic effect. 
Similar behavior has been reported in Ba(Fe$_{1-x}$Co$_x$)$_2$As$_2$, FeSe$_{1-x}$Te$_x$, and LiFeAs \cite{Gurevich2011}.
These distinct characteristics of $B_{\rm c2}$ between samples with $x = 0.12$ and 0.32 are most likely due to different pairing symmetries depending on the doping $x$.
\par The linear behavior for $x = 0.12$ and the concave upward feature for $x = 0.32$ cannot be understood by the single band model \cite{Werthamer1966}.
Hence, we employ the two-band model to analyze $B_{\rm c2}$.
The two-band $s$-wave model is based on strong interband couplings $\lambda_{12}$ and $\lambda_{21}$ together with repulsive intraband interactions $\lambda_{11}$ and $\lambda_{22}$ where 1 and 2 refer to either of electron or hole band.
For iron-based superconductors, no analysis for the upper critical field  has been made with the repulsive case $\lambda_{ii}<0$ \cite{Hunte2008, Jaroszynski2008, Kohama2009, Gurevich2011}.
In this study, we consider both cases of repulsive and attractive interactions utilizing the $B_{\rm c2}$ data taken throughout the whole temperature range below $T_{\rm c}$.
We use the phenomenological BCS-like theory considering neither retardation (dynamical) effect for the coupling constants, nor energy-dependent renormalization of quasi-particles.  Thus, we adopt the density of states derived by first-principle band calculations, and examine the sensitivity of the results on these values \cite{SM}. 
\par We first consider only the orbital effect for the pair breaking.
Another pair-breaking effect from the Pauli paramagnetism is included later. Then the temperature dependence of $B_{\rm c2}$ in the dirty limit is expressed as \cite{Gurevich2003}
\begin{eqnarray}
	\left(\lambda_{0}+\lambda_{-}\right)\left[\ln{t} + U_1\right]\nonumber
	+ \left(\lambda_{0}-\lambda_{-}\right)\left[\ln{t} + U_2\right]\nonumber
	\\+ {2w}\left[\ln{t} + U_1\right]\left[\ln{t} + U_2\right] = 0,
\end{eqnarray}
where $\lambda_{0} = ({\lambda_{-}}^2+4\lambda_{12}\lambda_{21})^{1/2}$, $\lambda_{-} = \lambda_{11} - \lambda_{22}$, $w = \lambda_{11}\lambda_{22}-\lambda_{12}\lambda_{21}$, $U_1 = U(h)$, $U_2 = U(\eta h)$, and $t = T/T_{\rm c}$.
Here, we have introduced $U(x) = \Psi(1/2+x)-\Psi(1/2)$ with $\Psi(x)$ being the digamma function, $\eta = D_{2}/D_{1}$ with $D_{i}$ being the generalized diffusivity in band $i$ for $H \parallel ab$-plane, and $h = {\hbar}D_{1}B_{\rm c2}(T)/[2\phi_{0}k_{\rm B}T]$.
The parameter $w$ serves as a measure of  relative coupling strength: $w > 0$ indicates the dominance of intraband couplings, while $w < 0$ shows the dominance of interband couplings.
The diffusivities $D_{1}$ and $D_{2}$ determine such observed quantities as the residual resistivity $\rho_{\rm n}$ and, together with $\lambda_{ij}$, the slope $dB_{\rm c2}/dT|_{T\rightarrow T_{\rm c}}\equiv B_{\rm c2}'$ of the upper critical field just below $T_{\rm c}$.
We can invert the relations to represent $D_1$ and $D_2$ in terms of $\rho_{\rm n}$ and $B_{\rm c2}'$ as
\begin{eqnarray}
	D_{1} &=& A\left( \frac{8k_B\phi_0}{\pi^2\hbar |B_{\rm c2}^{\prime}|}N_{2}-\frac{\lambda_0-\lambda_{-}}{\lambda_0}\frac{1}{e^2\rho_{\rm n}} \right),\\
	D_{2} &=& A\left( -\frac{8k_B\phi_0}{\pi^2\hbar |B_{\rm c2}^{\prime}|}N_{1}+\frac{\lambda_0+\lambda_{-}}{\lambda_0}\frac{1}{e^2\rho_{\rm n}} \right),
\end{eqnarray}
with $A = \lambda_0/[(\lambda_0 + \lambda_{-})N_2 - (\lambda_0 - \lambda_{-})N_1]$ and $e$ the elementary electric charge \cite{Gurevich2003}.
Here, $N_{1}$ and $N_{2}$ denote the partial densities of states on Fermi surfaces for the electron ($N_{\rm e}$) and hole ($N_{\rm h}$) bands, the values of which are obtained by {\it ab initio} calculations (see details in Supplementary Materials \cite{SM}).
We estimate $\rho_{\rm n}$ by extrapolating $\rho_{\rm n}(T)$ just above $T_{\rm c}$, as shown in Figs.~2(a) and 2(b).
\par
We now examine the temperature dependence of $B_{\rm c2}$ on coupling constants $\lambda_{ij}$ according to Eqs.~(1)--(3).
We employ the two-band $s$-wave model which is often used for iron-based superconductors.
In the model, the interband coupling should be stronger than the repulsive intraband coupling $\lambda_{11}\  (< 0)$ and $\lambda_{22}\ (< 0)$, which necessitates $w < 0$.
The transition temperature $T_c$ is determined by a combination of the dimensionless coupling constants $\lambda_{ij}$ and the cutoff energy $\hbar\omega_{\rm C}$, which corresponds to the Debye energy in the BCS theory.
The resultant scaling degrees of freedom prevents $\hbar\omega_{\rm C}$ from unique fixing for a given $T_c$.
We assume analogous scaling in magnetic field as well, as detailed in Supplementary Material [27].
Thus, we set $\hbar\omega_{\rm C}=1$ eV and search for $\lambda_{ij}$ so as to best fit $B_{\rm c2}(T)$.
Note that the interband couplings are characterized by a single parameter because of the constraint
$N_{1}\lambda_{12} = N_{2}\lambda_{21}$.
With another constraint to reproduce $T_c$, 
we left with two independent parameters for fitting, which are chosen as $\lambda_{11}$ and $\lambda_{22}$. 
%
\begin{center}
\begin{table*}
 \caption{Experimental results and fitting parameters based on the two-band theory. 
 Details of the fitting procedure with  $\hbar\omega_{\rm C}$ = 1 eV are explained in \cite{SM}. 
The symbols b1 and b2 denote bands-1 and -2, while $el$ and $h$ indicate electron and hole bands, respectively.
 Other quantitities such as $\eta$ and $D_0$ are defined in the main text. 
 $D_{1,0}$ has a unit of 10$^{-5}$m$^2$/s. $B_{\rm c2}(0)$ and $\Delta B_{\rm c2}$ have a unit of T. 
 }\label{tab:para_table}
 \small
	\begin{tabular}{cccccccccccccc}\\ \hline
	$x$ &b1&b2& $\lambda_{11}$ & $\lambda_{22}$ & $\left|\lambda_{12}\right|$ & $\left|\lambda_{21}\right|$ & $\eta$ & $D_1$ & $D_0$ & $w$ & $\lambda_{-}$ & $B_{\rm c2}(0)$ & $\Delta B_{\rm c2}$ \\ \hline
	0.12    & $el$ & $h$ & $-1$ & $-0.0563(19)$ & $ 0.566(2)$ & $0.430(2)$ & $0.057(1)$ & 32.9(1) & -- & $-$0.1872(3) & $-$0.9437(19) & 43.6(6) & 0.70\\
	0.12    & $h$ & $el$ & $-1$ & $-0.157(2)$ & $ 0.523(1) $ & $0.688(2)$ & $0.059(1)$ & 25.74(6) & -- & $-$0.2027(3) & $-$0.843(2) & 44.4(6) & 0.95\\
	0.12    & $el$ & $h$ & $1$ & $0.3359(3)$ & $ 0.4489(5) $ & $0.3416(4)$ & $8.91(3)$ & 2.782(8) & -- & 0.18259(5) & 0.6641(3) & 41.7(6) & 0.61\\
	0.12    & $h$ & $el$ & $1$ & $0.2809(2)$ & $ 0.2851(3) $ & $0.3746(4)$ & $10.61(3)$ & 2.964(8) & -- & 0.17408(3) & 0.7191(2) & 41.9(6) & 0.65\\
	0.32    & $h$ & $el$ & $0.165445(2)$ & $0.164492(1)$ & $ 0.0012(3) $ & $0.0008(2)$ & $0.3(1)$ & 7(1) & 2.09(1) & 0.0272133(4) & 0.000953(2) & 75(13) & 0.63\\
	\hline
	\end{tabular}
\end{table*}
\end{center}
\par For $x=0.12$, the fitting reproduces the experimental data for a wide range of $\lambda_{11}$ and $\lambda_{22}$ as explained in Supplementary Material \cite{SM}.
Here, we set $\lambda_{11} = \pm 1$ for simplicity, and search for other parameters $\lambda_{ij}$ to best fit $B_{\rm c2}(T)$.
Table~I summarizes the results for the parameters obtained.
We obtain good fits whether the source of band-1 is electron or hole. 
The red solid (dashed) line in Fig. 4 depicts the typical fitting in the scenario when band-1 corresponds to the hole band and $\lambda_{11} = -1$ (1). 
For $x=0.12$, reasonable results are obtained for 
a case of repulsive intraband interaction $\lambda_{11}=-1$, and another case of 
attractive interaction $\lambda_{11}=1$.
The other parameters in the fit are listed in Table I.
The repulsive case with dominant interband coupling, as indicated by $w<0$, fits the $s$-wave paradigm of iron-based superconductor.
In the attractive case with $\lambda_{11} = 1$, on the other hand, the interband coupling indeed helps the pairing, but is not dominant as indicated by $w>0$ in Table I .
In any case, the origin of attraction is hard to be identified in the presence of strong Coulomb repulsion \cite{Chubukov2012}.
Thus, we favor the case of repulsive intraband interaction for the SC1 superconductivity.
\par In contrast, for $x = 0.32$, Eq. (1) fails to reproduce the data for any choice of parameters.
We thus include the Pauli paramagnetic effect alongside the orbital effect.
Namely we replace $U_n\ (n=1,2)$ by $U_n^*$ as follows: \cite{Gurevich2007, Jaroszynski2008}
\begin{equation}
U^{\ast}_n = {\rm Re}\Psi\left[ 1/2 + l(D_n /  D_0 + i) \right] - \Psi(1/2),
\end{equation}
with $l = {\hbar}D_0B_{\rm c2}(T) /[ 2\phi_0 k_{\rm B} T]$ and $D_{0} = g \mu_{\rm B}\phi_{0}/2\sqrt{2}{\pi}\hbar$.
Here, $\mu_{\rm B}$ is the Bohr magneton, and $g$ is the effective $g$-factor.
The Zeeman effect represented by  $D_0$  realizes the Pauli limit.
Only with the condition $w > 0$ and $\lambda_{ii} >0$,  we can determine the three parameters of $\lambda_{11}$, $\lambda_{22}$, and $D_0$ by the least squares method and can reproduce the measured $B_{\rm c2}(T)$ reasonably well.
The estimated values with preset $\hbar\omega_{\rm C}=1$ eV are listed in Table I.
\par Here, we have assigned hole as band-1 source and electron as band-2 source.
For easy comparison the results with and without the Pauli paramagnetic effect are shown in Fig.~4 by the blue solid and dotted lines, respectively.
Note that $B_{\rm c2}'$ is not affected by the Pauli paramagnetic effect.
The concave upward trend toward $T_{\rm c}$ exhibited in Fig.~4  also suggests the presence of two weakly linked bands with $w>0$.
\par Although we have assumed the $s$-wave pairing in the analysis, the strong Coulomb repulsion favors only the interband mechanism for the $s$-wave pairing,
which is at odds with fitted couplings at $x=0.32$.
In the case of cuprates, the attractive interaction in a single band is associated with the $d$-wave pairing. 
It has been suggested for heavily electron-doped iron-based superconductors that the $d$-wave paring channel promotes superconductivity \cite{Chubukov2012}.
In fact, LaFeAsO$_{1-x}$H$_{x}$ in the higher $x$-region exhibits anomalous behavior in resistivity \cite{Iimura2016}, large magnetic moment \cite{Hiraishi2014}, and gapless magnetic excitation, which seems to stem from single orbital character \cite{Tamatsukuri2018}. 
Currently, however, there is no practical theory to analyze $B_{\rm c2}$ for non $s$-wave pairings with disorder.  This is in strong contrast
with the $s$-wave pairing which enjoys much simpler theoretical scheme in the dirty limit  \cite{Werthamer1966}. 
\par
We here discuss the origin of pair breaking at $B_{\rm c2}$.
In many iron-based superconductors, $B_{\rm c2}$ surpasses the Pauli-limited $B_{\rm P}^{\rm BCS}$in the BCS theory \cite{Gurevich2011}.
The Pauli limit in the BCS theory is given by $B_{\rm P}^{\rm BCS}(0) = \sqrt{2} \Delta _0 /(g\mu _{\rm B})$, where $\Delta _0$ is the superconducting gap at 0 K.
The above equation yields the well-known relation $B_{\rm P}^{\rm BCS}(0)=1.86~T_{\rm c}$ in a weak-coupling superconductor with $g=2$.
The values of $B_{\rm P}^{\rm BCS}(0)$ for $x = 0.12$ and 0.32 are estimated to be 37.9 and 54.7~T, respectively, which are lower than the current results of $B_{\rm c2}(0)$ = 43.2 and 77.4~T.
We infer that the difference between $B_{\rm P}^{\rm BCS}(0)$ and experimental $B_{\rm c2}(0)$ comes from the strong-coupling  which raises $\Delta _0/T_c$, and/or the spin-orbit coupling which decreases $g$.
\par Inelastic neutron scattering has determined the superconducting gap for $x$ = 0.10 and 0.35 as $\Delta _0$ = 7.8 and 7.0 meV \cite{Yamaura2019,Hiraka2020}, implying a strong coupling superconductor with 2$\Delta_0/k_{\rm B}T_{\rm c}$ = 7.0 and 4.5, respectively.
When we use measured $\Delta _0$ to incorporate the strong coupling effect in $B_{\rm P}^{\rm BCS}(0)$, we get $B_{\rm P}^{\ast}(0)$ = 95.3~T for $x = 0.10$, and 85.5~T for $x = 0.35$.
Comparing $B_{\rm P}^{\ast}(0)$ with $B_{\rm c2}(0)$ for $x = 0.12$ and 0.32, $B_{\rm P}^{\ast}(0)$ is substantially higher than our result in SC1, but is comparable to that in SC2.
In other words, the paramagnetic effect becomes dominant in SC2 when the orbital limit significantly exceeds the Pauli limit.
This situation is analogous to the upper critical field of cuprate superconductors with extremely short coherence length \cite{Nakamura2019}.
Therefore, we conclude that the orbital effect in SC1 and the paramagnetic effect in SC2 are primarily responsible for pair breakings at $B_{\rm c2}$.
Since our two-band model is based on the weak-coupling theory, it is desirable to improve the model to take account of the strong-coupling effect more precisely.
\par In summary, we find the peculiar doping dependence of the upper critical field for LaFeAsO$_{1-x}$H$_x$ with two superconducting phases, SC1 ($x$ = 0.12) and SC2 ($x$ = 0.32).
By the two-band analysis assuming the $s$-wave pairing, we have shown that the superconductivity in SC1 is consistent with dominant interband coupling in the presence of repulsive intraband interaction.
In SC2, however, the attractive intraband interaction is dominating with weak interband couplings.
Then another superconducting mechanism, such as the $d$-wave model, should be explored in SC2.
Moreover, we find that pair breakings at $B_{\rm c2}$ dominantly come from the orbital effect in SC1, whereas from the paramagnetic effect in SC2. 
\begin{acknowledgments}
This work was supported by the MEXT Elements Strategy Initiative to Form Core Research Center (JPMXP0112101001) and JSPS KAKENHI (No. 16K05434). This work was carried out by the joint research in the Institute for Solid State Physics, the University of Tokyo.
\end{acknowledgments}


\begin{thebibliography}{99}
\bibitem{Hiraka2020}
The superconducting gap $\Delta_0$ = 7.0~meV for LaFeAsO$_{0.65}$D$_{0.35}$ was determined by inelastic neutron scattering as well as Ref.~34 (unpublised).
\end{thebibliography}
\end{document}